\documentclass[aps,reprint,amsmath,amssymb,superscriptaddress]{revtex4-2}
\usepackage{graphicx}
\usepackage{float}
\usepackage{dcolumn}
\usepackage{bm}
\begin{document}
\title{Schrödinger-Heisenberg Variational Quantum Algorithms}
\author{Zhong-xia Shang}
\affiliation{Hefei National Laboratory for Physical Sciences at Microscale and Department of Modern Physics, University of Science and Technology of China, Hefei, Anhui 230026, China}
\affiliation{Shanghai Branch, CAS Centre for Excellence and Synergetic Innovation Centre in Quantum Information and Quantum Physics, University of Science and Technology of China, Shanghai 201315, China}
\affiliation{Shanghai Research Center for Quantum Sciences, Shanghai 201315, China}
\author{Ming-cheng Chen}
\affiliation{Hefei National Laboratory for Physical Sciences at Microscale and Department of Modern Physics, University of Science and Technology of China, Hefei, Anhui 230026, China}
\affiliation{Shanghai Branch, CAS Centre for Excellence and Synergetic Innovation Centre in Quantum Information and Quantum Physics, University of Science and Technology of China, Shanghai 201315, China}
\affiliation{Shanghai Research Center for Quantum Sciences, Shanghai 201315, China}
\author{Xiao Yuan}
\affiliation{Center on Frontiers of Computing Studies, Peking University, Beijing 100871, China}
\affiliation{School of Computer Science, Peking University, Beijing 100871, China}
\author{Chao-yang Lu}
\affiliation{Hefei National Laboratory for Physical Sciences at Microscale and Department of Modern Physics, University of Science and Technology of China, Hefei, Anhui 230026, China}
\affiliation{Shanghai Branch, CAS Centre for Excellence and Synergetic Innovation Centre in Quantum Information and Quantum Physics, University of Science and Technology of China, Shanghai 201315, China}
\affiliation{Shanghai Research Center for Quantum Sciences, Shanghai 201315, China}
\author{Jian-wei Pan}
\affiliation{Hefei National Laboratory for Physical Sciences at Microscale and Department of Modern Physics, University of Science and Technology of China, Hefei, Anhui 230026, China}
\affiliation{Shanghai Branch, CAS Centre for Excellence and Synergetic Innovation Centre in Quantum Information and Quantum Physics, University of Science and Technology of China, Shanghai 201315, China}
\affiliation{Shanghai Research Center for Quantum Sciences, Shanghai 201315, China}
\begin{abstract}
Recent breakthroughs have opened the possibility to intermediate-scale quantum computing with tens to hundreds of qubits, and shown the potential for solving classical challenging problems, such as in chemistry and condensed matter physics. However, the high accuracy needed to surpass classical computers poses a critical demand to the circuit depth, which is severely limited by the non-negligible gate infidelity, currently around $0.1-1\%$. The limited circuit depth places restrictions on the performance of variational quantum algorithms (VQA) and prevents VQAs to explore desired non-trivial quantum states. To resolve this problem, we propose a paradigm of Schrödinger-Heisenberg variational quantum algorithms (SH-VQA). Using SH-VQA, the expectation values of operators on states that require very deep circuits to prepare can now be efficiently measured by rather shallow circuits. The idea is to incorporate a virtual Heisenberg circuit, which acts effectively on the measurement observables, to a real shallow Schrödinger circuit, which is implemented realistically on the quantum hardware. We choose a Clifford virtual circuit, whose effect on the Hamiltonian can be seen as an efficient classical processing. Yet, it greatly enlarges the state expressivity, realizing much larger unitary $t$-designs. Our method enables accurate quantum simulation and computation that otherwise are only achievable with much deeper circuits or more accurate operations conventionally. This has been verified in our numerical experiments for a better approximation of random states, higher-fidelity solutions to the XXZ model, and the electronic structure Hamiltonians of small molecules. Thus, together with effective quantum error mitigation, our work paves the way for realizing accurate quantum computing algorithms with near-term quantum devices.
\end{abstract}
\maketitle

\begin{figure}[htb]
\includegraphics[width=0.42\textwidth]{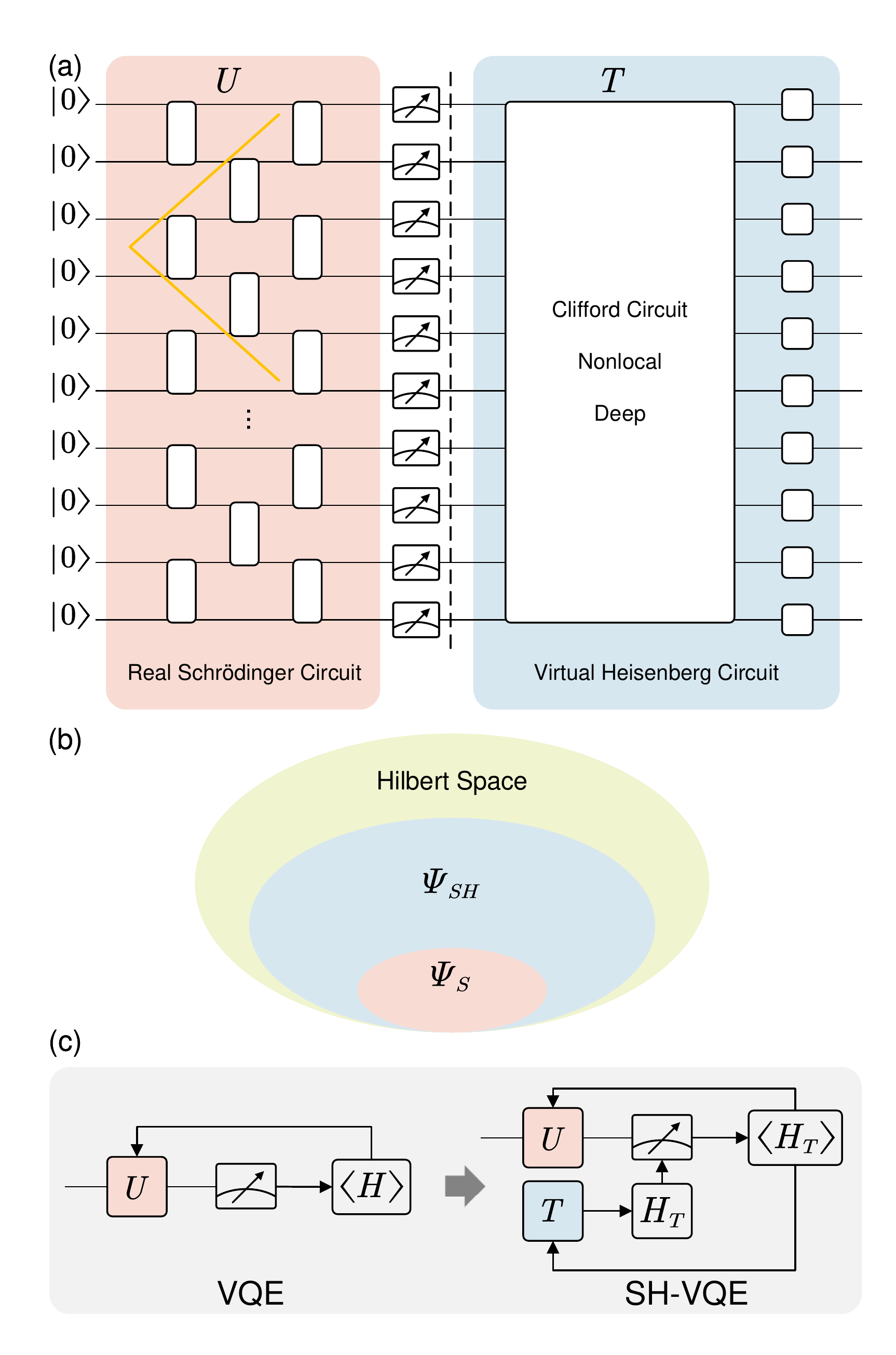}
\caption{\label{p1} SH-VQE. (a): The SH-VQE circuit. The circuit is composed of the Schrödinger circuit $U$ and the Heisenberg circuit $T$, where $U$ is the local unitary circuit running on real quantum computers and $T$ is the virtual circuit acted on the Hamiltonian consisting of two parts, the Clifford part, and the single qubit layer. The architecture we use for $U$ throughout this work is layers of parallel 2-qubit gates, which has a well-defined light cone that constrains the propagation of correlations and entanglements. (b): Improvements of SH-VQE. By adding the virtual circuit, $T U\left|0^{\otimes n}\right\rangle$ is able to explore more of the Hilbert space compared with $U\left|0^{\otimes n}\right\rangle$ in conventional VQE and the trainable Hilbert space is much larger than the conventional VQE. (c): Algorithm structure comparison between VQE and SH-VQE. The transformed Hamiltonian $H_{T}$ replaces $H$ in SH-VQE. We update parameters in both $U$ and $T$ to minimize the expectation value of $H_{T}$.}
\end{figure}

Almost four decades after Richard Feynman put forward the idea of quantum computing \cite{feynman2018simulating}, the quantum advantage has been experimentally tested recently in the solid state systems\cite{arute2019quantum,zhu2022quantum,wu2021strong} and photonic systems \cite{zhong2020quantum,zhong2021phase}. However, those quantum computational advantage works focused on well-defined quantum sampling problems which were not designed practically useful. Therefore, the next important near-term milestone is to find algorithms for noisy intermediate-scale quantum (NISQ) \cite{preskill2018quantum} devices to solve non-trivial practical problems that are intractable for classical computation.

One of the most promising NISQ applications is using variational quantum algorithms (VQA) \cite{endo2021hybrid,cerezo2021variational} such as the variational quantum eigensolver (VQE) \cite{peruzzo2014variational} and the variational quantum simulation (VQS) \cite{yuan2019theory} where a quantum circuit is optimized classically to approximate the eigenstate state energy and to simulate the dynamics of a Hamiltonian respectively for tasks that are widely considered in combinatorial optimization problems \cite{farhi2014quantum}, condensed matter physics \cite{wecker2015progress}, and quantum chemistry \cite{kandala2017hardware,romero2018strategies}. A practical advantage of hybrid algorithms is their certain degrees of resilience to noise in the optimization and quantum hardware \cite{endo2021hybrid,temme2017error,li2017efficient}.

Considering the limitations of NISQ devices, VQAs generally use a shallow local unitary circuit (LUC) (Fig. \ref{p1}(a)) to approximate the target quantum states. States prepared by shallow LUCs however, could be trivial, obeying the entanglement area law \cite{brandao2015exponential} which can be well captured by classical tensor networks \cite{verstraete2008matrix}. Indeed, the Lieb-Robinson bound \cite{bravyi2006lieb} indicates that the entanglement light cone restricts the propagation of correlations and, therefore, shallow LUC can not generate long-range entanglement. However, the ground states of some Hamiltonians of interest could be highly non-trivial and require a relatively deep LUC with a depth that has linear or even higher scaling with the qubit number \cite{bravyi2006lieb,ho2019efficient} such as interacting spins at critical points \cite{vidal2003entanglement,latorre2003ground}, topological quantum orders \cite{huang2015quantum,chen2010local}, and interacting fermions in complex molecules \cite{romero2018strategies}. This is a big challenge for NISQ devices. Indeed, without an effective quantum error correction, the final fidelity of the quantum circuits drops exponentially with the number of gates. For example, a state-of-the-art random quantum circuit with 60 qubits and 24 layers \cite{zhu2022quantum} ended up with a cross-entropy benchmarking fidelity as low as $0.037 \%$. We thus need to significantly improve the NISQ hardware to implement those VQAs to the desired accuracy.

This situation can be summarized as a tradeoff between the fidelity of the LUC and its expressivity \cite{cerezo2021variational} (i.e., the ability for the quantum circuits to ``express" a sufficiently large volume of quantum states to include those non-trivial ones). To circumvent this problem, we propose a new framework of VQAs, enhanced by virtual Heisenberg circuits, which can noiselessly increase the effective circuit depth and thus simultaneously improve its expressivity and fidelity. We want to mention that there is a related work by Zhang et al. where their classical neural networks serve for a similar purpose as our virtual Heisenberg circuits \cite{zhang2022variational}. And there is an orbital optimized unitary coupled cluster method \cite{mizukami2020orbital} that shares a similar idea as ours where they turn single-excitation circuits into a classical processing on chemical Hamiltonians. We call our scheme Schrödinger-Heisenberg (SH) variational quantum algorithms (VQA), which illustrates that the main idea is that, in addition to the physical unitary circuit, $U$, acting on the quantum states in the Schrödinger picture, we bring in a virtual circuit, $T$, acting on the target Hamiltonian $H$ in the Heisenberg picture (see Fig. \ref{p1}(a)). In the following, we consider SH-VQE as an example, but we note that the algorithm works for general VQAs. In this case, the energy expectation value $E(T, U)=$ $\left\langle 0^{\otimes n}\left|U^{\dagger} T^{\dagger} H T U\right| 0^{\otimes n}\right\rangle$ of the system becomes
\begin{equation}\label{1}
E(T, U)=\left\langle 0^{\otimes n}\left|U^{\dagger} H_{T} U\right| 0^{\otimes n}\right\rangle
\end{equation}
where the classically calculated transformed Hamiltonian $H_{T}=T^{\dagger} H T$ has the same energy spectrum as $H$. By properly choosing a relatively deep but noiseless $T$, the state $T U\left|0^{\otimes n}\right\rangle$ could explore the Hilbert space far outside the range of $U\left|0^{\otimes n}\right\rangle$ (see Fig. \ref{p1}(b)) and hence can obtain lower and more accurate ground-state energy than conventional VQE for non-trivial problems. We show a workflow of SH-VQE together with a comparison to conventional VQE in Fig. \ref{p1}(c). Compared with VQE, both the real Schrödinger circuit $U$ and the virtual Heisenberg circuit $T$ in SH-VQE are parametrized and updated when minimizing the expectation value $E(T, U)$. The key feature of SH-VQE is that only $U$ as a shallow LUC is physically implemented, whereas the relatively deep circuit $T$ is performed virtually and noiselessly using a classical computer.

We first show how to effectively measure $H_{T}$. In general, the target Hamiltonian $H$ could be expressed as a linear sum of multi-qubit Pauli terms $H=\sum_{i=1}^{m} g_{i} P_{i}$, where $P_{i} \in$ $\left\{\sigma_{I}, \sigma_{X}, \sigma_{Y}, \sigma_{Z}\right\}^{\otimes n}$. Then we can measure each $P_{i}$ with a total number of samples $\left(\frac{m}{\epsilon^{2}}\right) \sum_{i}$ $g_{i}^{2} \operatorname{Var}\left[P_{i}\right]$, proportional to the number of terms $m$ in the Hamiltonian \cite{mcclean2016theory}, to evaluate the energy expectation value within an error of $\epsilon$. Here $\operatorname{Var}\left[P_{i}\right]=\left\langle P_{i}^{2}\right\rangle-\left\langle P_{i}\right\rangle^{2}$. We can similarly measure $H_{T}$, by similarly decomposing each $T^{\dagger} P_{i} T$ into Pauli strings. While most practical Hamiltonians $H$ only contain a polynomial number of terms, this might not be the case for $T^{\dagger} P_{i} T$ or $H_{T}$, after the transformation (See Appendix).

\begin{figure}[htb]
\includegraphics[width=0.49\textwidth]{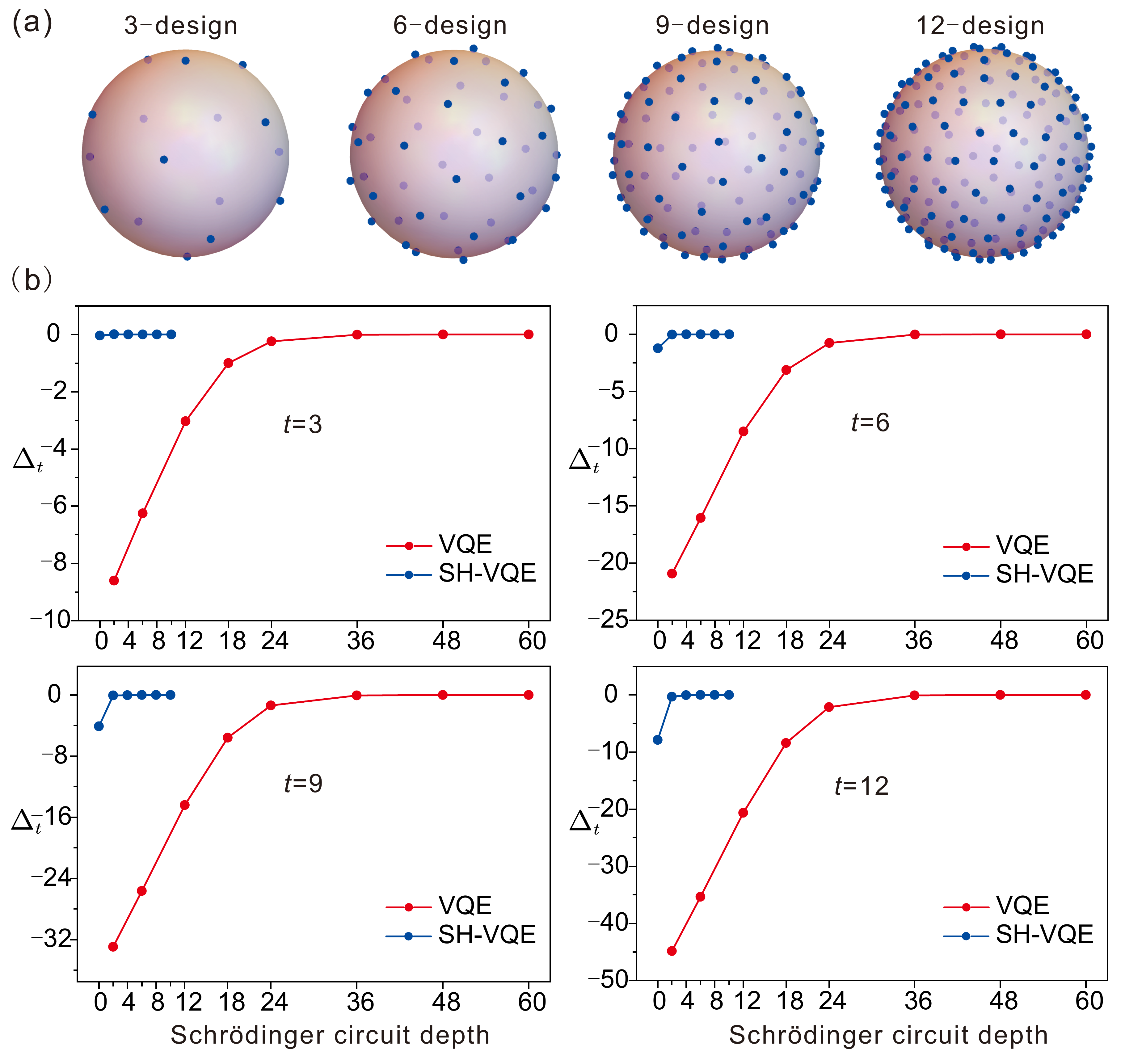}
\caption{\label{p2} SH-VQE expressivity. (a): Relationship between expressivity and the $t$-design. We show the point distribution on the Bloch sphere of different design orders $t=3,6,9,12$. (b): Comparison of the expressivity measure $\Delta_{t}$ between VQE and SH-VQE. The structure of the Clifford part is formed by 500 randomly picked basic Clifford gates in the set $\{\mathrm{H}, \mathrm{S}, \mathrm{CNOT}\}$. The other parts including the two-qubit blocks in Schrödinger LUC and gates in the Heisenberg single qubit layer are random gates drawn from the Haar measure. The zero-depth setting in SH-VQE can be understood as the performance of the Clifford circuit. Since Clifford circuit can generate 3-design, $\Delta_{t}$ approaches 0 for $t=3$ whereas below 0 in other cases. Schrödinger circuit of depth greater than 6 combined with the Heisenberg circuit is believed to generate the maximally scrambled states since values of $\Delta_{t}$ from $t=3$ to $t=$ 12 are all zero \cite{liu2018entanglement}.}
\end{figure}

Here we propose a structure of the Heisenberg circuit that also leads to efficiently measurable $T^{\dagger} P_{i} T$ or $H_{T}$. The circuit consists of two parts (Fig. \ref{p1}(a)), where the first part is an arbitrary Clifford circuit that can be decomposed into a sequence of $O\left(n^{2}\right)$ basic gates from the set $\{\mathrm{H}, \mathrm{S}, \mathrm{CNOT}\}$, and the second part is a layer of single-qubit gates. The first part realizes discrete gates such as CNOT to build correlations between any two qubits and the second part makes them continuous. The Clifford circuit maps the multi-qubit Pauli group to itself, which conserves the number of terms of the Hamiltonian. Also, the Gottesman-Knill theorem \cite{gottesman1998heisenberg} indicates that calculating the transformed Hamiltonian is easy. While the second part might increase the number of terms of the Hamiltonian, the overhead is polynomial for Hamiltonians $H$ consisting of only $k$-weight terms, i.e., the Pauli operators $\left\{\sigma_{X}, \sigma_{Y}, \sigma_{Z}\right\}$ act on at most $k$ qubits since the weight remains unchanged. We note that one can change this part into other easier or more complex circuits for different Hamiltonians, considering the trade-off between the circuit power and the measurement cost. 

We begin to study the expressivity of the circuit in SH-VQE. We consider the expressivity measure using the method of quantum complex projective $t$-design \cite{hoggar1982t}, which means that the distribution of the output states has equal moments up to the $t^{\text {th }}$ order to a Haar uniform distributed states from the whole Hilbert space. Intuitively, as illustrated in Fig. \ref{p2}(a) \cite{Note2}, a higher $t$-design indicates a more uniform and denser state distribution in the Hilbert space, and vice versa. In general, a LUC of depth $O\left(n t^{10}\right)$ is needed to generate a $t$-design \cite{brandao2016local}, and the Clifford circuits can produce a 3-design \cite{zhu2017multiqubit}. Using the tight Page's theorem \cite{liu2018entanglement}, we define the logarithmic difference of entanglement entropy as
\begin{equation}\label{2}
\Delta_{t}=\log \left(E_{\text {Haar }}\left[\operatorname{Tr}\left(\rho_{n / 2}^{t}\right)\right]\right)-\log \left(E_{\mathrm{SH}}\left[\operatorname{Tr}\left(\rho_{n / 2}^{t}\right)\right]\right)
\end{equation}
to identify the order of expressivity of SH-VQE, where $\rho_{n / 2}$ is the reduced half system density matrix, $E_{\text {Haar }}$ is the average over Haar random states, and $E_{S H}$ is the average over the quantum states $T U\left|0^{\otimes n}\right\rangle$. If $\Delta_{t}$ increases and approaches 0 , it means that $T U\left|0^{\otimes n}\right\rangle$ is a $t$-design.

Fig. \ref{p2}(b) shows a comparison of the expressivity of SH-VQE versus VQE through a numerical experiment on a 12-qubit system. In the VQE setting, we run a random LUC at different depths and calculate $\Delta_{t}$ to characterize the $t$-design. In the SH-VQE setting, we implement both the real Schrödinger circuits $U$ and the virtual Heisenberg circuits $T$ which are pure Clifford consisting of 500 random gates from $\{\mathrm{H}, \mathrm{S}, \mathrm{CNOT}\}$. The key observation for both cases is the critical depths when the $\Delta_{t}$ measure increases to and saturates at around 0. It is evident that the $\Delta_{t}$ curves for SH-VQE rise much more rapidly than that for VQE for all $t$ values from 3 to 12. The rising curve for SH-VQE quickly hits the saturation point at a Schrödinger circuit depth of $\sim 2$, while the VQE curve arrives at a much deep depth of $\sim 36$. This indicates that SH-VQE can effectively reduce the gate depth by more than one order of magnitude to achieve the same level of expressivity. For a higher number of qubits, we expect an even more dramatic advantage, which can be inferred from a qubit-size dependent test of depth reduction as shown in Fig. \ref{ps3}. The above results indicate that we can use current NISQ hardware to effectively run deep quantum circuits while maintaining high fidelity. Particularly, based on a two-qubit gate fidelity of $99.5\%$, the SH-VQE can allow us to run, for instance, a 12-qubit 4-depth quantum circuit with an output fidelity of $90 \%$, which would otherwise demand a two-qubit gate fidelity of $99.95 \%$ (currently unrealistic) and depth of 40 in conventional VQE (Fig. \ref{ps2}). Note that shallow LUCs or Clifford circuits alone can only generate small design orders, but a combination of them can achieve high expressivity.
    
We consider an example of the XXZ spin model with a periodic boundary condition
\begin{equation}\label{3}
H_{X X Z}=\sum_{i=1}^{n}\left[\sigma_{i}^{x} \sigma_{i+1}^{x}+\sigma_{i}^{y} \sigma_{i+1}^{y}+\Delta \sigma_{i}^{z} \sigma_{i+1}^{z}\right]
\end{equation}
to demonstrate a kind of working flow of SH-VQE. At the critical point $\Delta=1$, the $\mathrm{XXZ}$ model is equivalent to the Heisenberg model whose ground state has a logarithmic scaling of entanglement entropy \cite{vidal2003entanglement,latorre2003ground}, and hence cannot be prepared by a constant-depth LUC. Since we aim to boost the performance of the NISQ experiments, we use the hardware efficient ansatz \cite{kandala2017hardware} for the real Schrödinger circuit even though this may lead to barren plateau problems \cite{mcclean2018barren}, where each circuit layer composes a layer of CZ gates and a layer of parametrized arbitrary single-qubit gates (denoted as $\vec{\theta})$

\begin{figure*}[htb]
\includegraphics[width=0.8\textwidth]{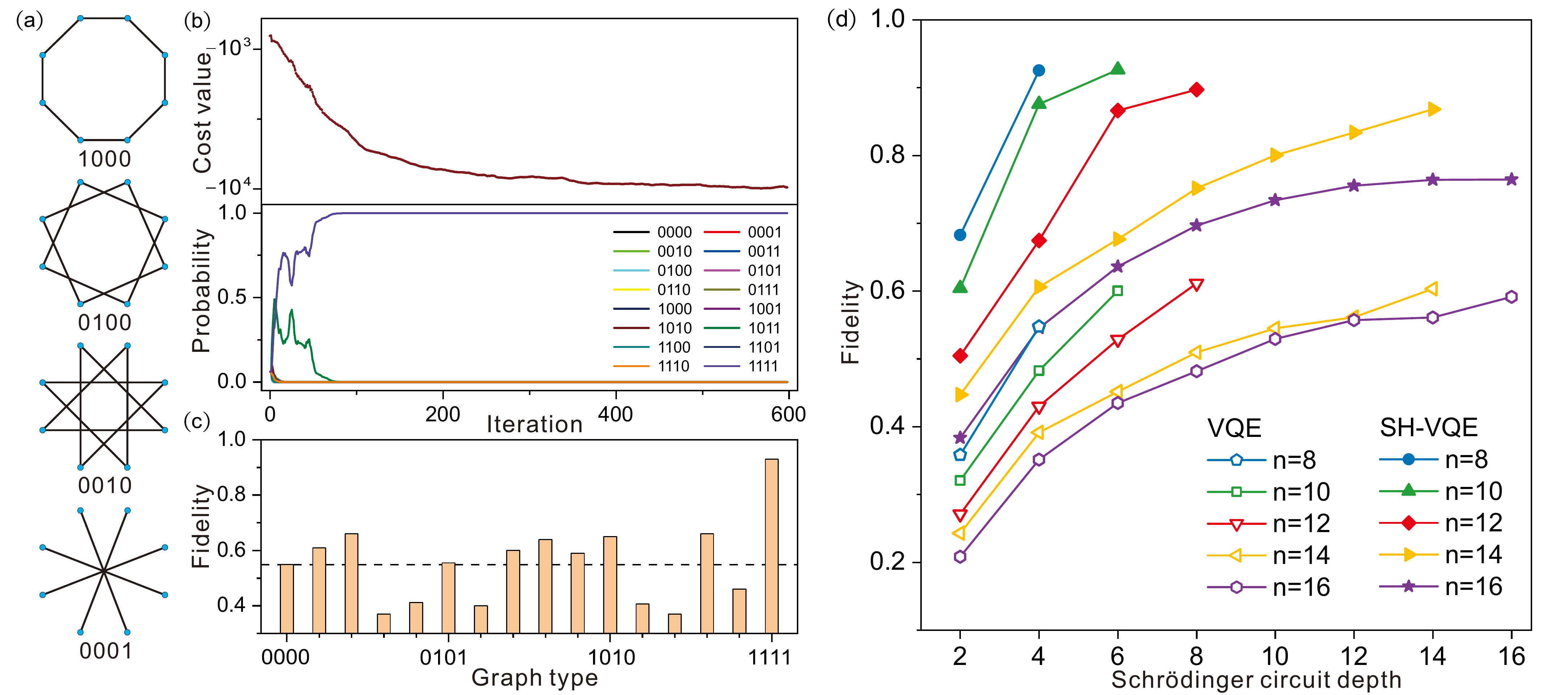}
\caption{\label{p3} Searching the Clifford circuit for the XXZ model. (a): The 4 elementary graphs and their corresponding code strings for $n=8$ TI graphs. (b): Upper Panel: Minimizing the cost function to search for the best graph. Parameters contain both gate parameters and probability parameters. The cost function is the sum of the Hamiltonian expectation values of circuits sampled from $\vec{\alpha}$ under the same gate parameters. The number of samples at each iteration is 800. Lower Panel: Probabilities of all 16 graphs as functions of iteration times. All graphs have the same probabilities at the beginning. The probability of the fully connected graph `1111' becomes 1 as the iteration times grow. The optimization algorithm used for circuit structure searching is adam-SPSA \cite{arouri2020accelerated}. (c): Direct comparisons between different graphs. The dashed line is the result of the graph type: `0000' i.e. without the Heisenberg circuit. The fully connected graph is indeed the best choice. There exist graphs that have worse performance than `0000'. (d): Comparison of solved ground state fidelities. We use VQE and SH-VQE to solve 8, 10, 12,14, and 16-qubit XXZ models. Fully connected graphs are used as the Clifford layer. VQE and SH-VQE of the same Schrödinger circuit depth share the same color. Each point is the best result obtained from 20 sets of random initial parameters.}
\end{figure*}

\begin{figure}[htb]
\includegraphics[width=0.49\textwidth]{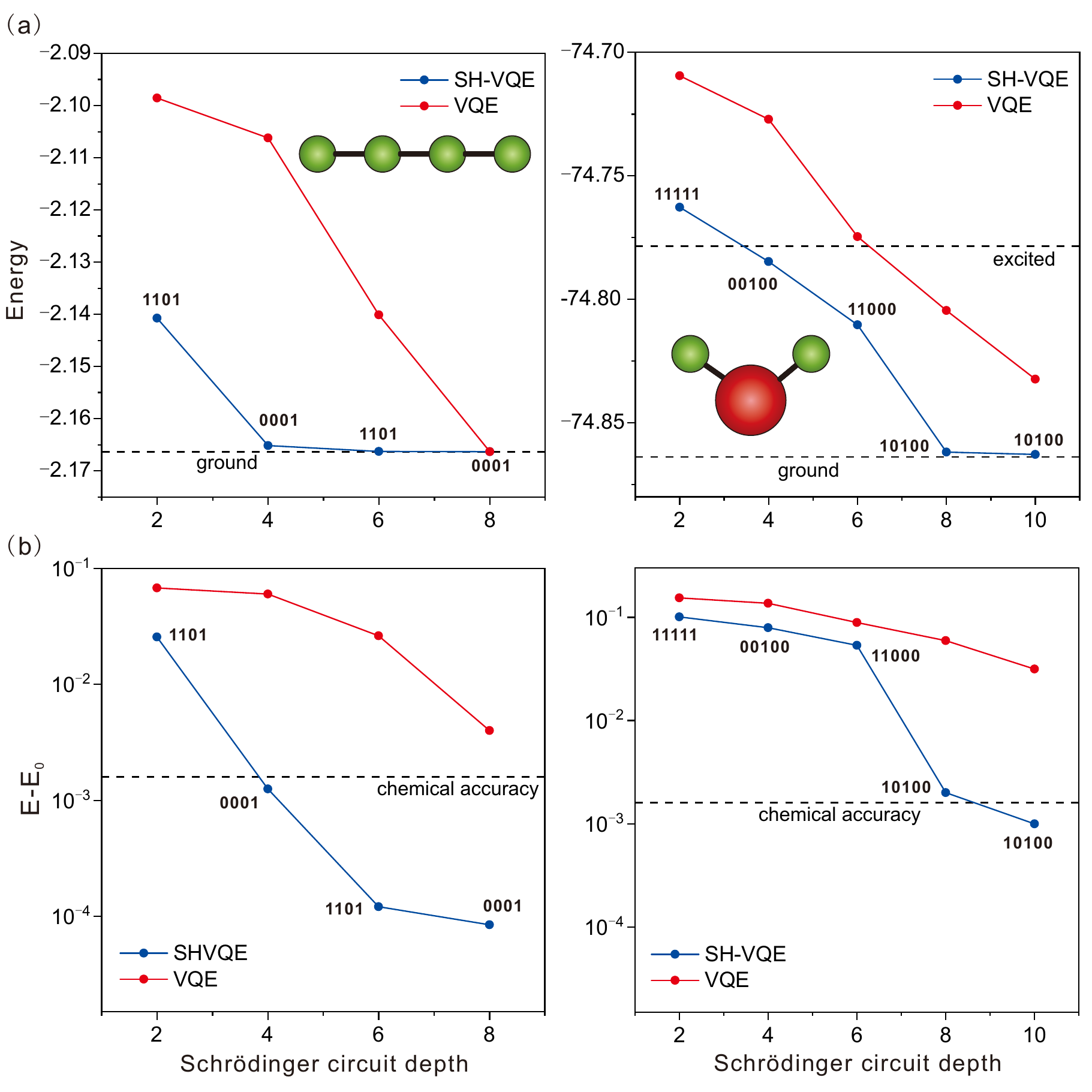}
\caption{\label{p5} Comparisons of SH-VQE and VQE on solving small molecules. Using VQE and SH-VQE to solve the 8-qubit Hamiltonian of the $\mathrm{H}_{4}$ molecule of the bond distance 1.0 Angstrom and the 10-qubit Hamiltonian of the $\mathrm{H}_{2} \mathrm{O}$ molecule of the bond distance 1.5 Angstrom. The binary string around the SH-VQE data label the searched optimal graph circuit. (a): Solved energy as a function of Schrödinger circuit depth. (b): Absolute energy differences as functions of Schrödinger circuit depth.}
\end{figure}

For the Heisenberg circuit, the single-qubit gate layer is parametrized with parameters $\vec{\phi}$. And we restrict the Clifford part to graph circuits \cite{hein2006entanglement} where only commuting $\mathrm{CZ}$ gates are used. We separate the graph circuit into patterns of different connectivity with the same translational invariant (TI) symmetry as $H_{X X Z}$. More concretely, for an $n$-qubit circuit, we can set $\lfloor n / 2\rfloor$ elementary graphs (For the $\mathrm{j}^{\text {th }}$ elementary graph, each node $\mathrm{i}$ is connected with node $i+j$. $\lfloor\cdot\rfloor $ is the floor function.). As each elementary graph can be turned on or turned off, the total number of possible patterns is $2^{\lfloor n / 2\rfloor}$ and we use a $\lfloor n / 2\rfloor$-bit string to label all the possible patterns such as `$01001 \ldots$', where 0 means the corresponding elementary graph is turned on whereas 1 means off (Fig. \ref{p3}a). To efficiently search through an exponentially large space of Clifford gate patterns, we borrow the idea from differentiable quantum architecture search \cite{zhang2020differentiable}, where each elementary TI graph is turned on independently according to a probability described by a two-parameter softmax function \cite{Note1}. Thus, only $\lfloor n / 2\rfloor \times 2$ parameters (denoted as $\vec{\alpha}$ ) are needed to implement the discrete search of the huge Clifford patterns. Therefore, the circuit ansatz for the SH-VQE is
\begin{equation}\label{4}
T(\vec{\alpha}, \vec{\phi}) U(\vec{\theta})\left|0^{\otimes n}\right\rangle
\end{equation}
where $\vec{\alpha}$ and $\vec{\phi}$ represent all configurations of the Heisenberg circuit $T$ and $\vec{\theta}$ are the continuous parameters in the single-qubit gates inside the Schrödinger circuit $U$. The parameters $\vec{\alpha}$ are used to generate samples of different circuits and the cost function is the average of the Hamiltonian expectation values of these circuits under the same gate parameters $\vec{\theta}$ and $\vec{\phi}$. The SH-VQE method then optimizes over all the parameters to search for the ground state of the Hamiltonian.  

In our numerical simulation, we consider an 8-spin XXZ model with a 4-depth circuit $U$ and 4 elementary TI graphs of the Clifford layer as shown in Fig. \ref{p3}(a). We show the energy expectation and the evolution of the possibilities of all 16 configurations during the optimization as functions of the number of iterations in Fig. \ref{p3}(b). When the energy expectation is converged, the probabilities of the candidate circuit structures concentrate on the optimal configuration, the fully connected graph `1111'. In Fig. \ref{p3}(c), we show the optimal energies of all the 16 candidate circuit configurations, which verifies that the optimal configuration is indeed the fully connected graph `1111'. We further solve larger models up to 16 spins to show the improvement of SH-VQE compared with conventional VQE using the same Schrödinger circuits (Fig. \ref{p3}(d)). For SH-VQE, we directly use the generalized fully connected graph circuits as the Clifford part. We can find under the same circuit depth, the SH-VQE obtains higher fidelities than the VQE (an average improvement of $25.2 \%$).

To further demonstrate the practical values of our algorithm, we implement our algorithm to solve the electronic structure problems of $\mathrm{H}_{4}$ and $\mathrm{H}_{2} \mathrm{O}$ molecules following the same workflow as the above. The $\mathrm{H}_{4}$ molecule corresponds to an 8-qubit Hamiltonian. For the $\mathrm{H}_{2} \mathrm{O}$ molecule, we use the active space method \cite{mcardle2020quantum} to create an effective 10-qubit Hamiltonian containing 10 spin orbitals and 6 electrons. Since the SH-VQA has the Pauli weight restriction, we use the Bravyi-Kitaev mapping which transforms an $M$-mode fermionic Hamiltonian to a spin Hamiltonian of $O\left(\log _{2} M\right)$ Pauli weight \cite{bravyi2002fermionic,mcardle2020quantum}. Note that the ground states of these molecule Hamiltonians have the correct number of electrons. The results are shown in Fig. \ref{p5}, where we can see SH-VQE can reach the chemical accuracy ($1.6\times 10^{-3}$) with Schrödinger circuits of much shallower depth than VQE.

We now give some discussions for SH-VQA. First, we want to emphasize that the states $T U\left|0^{\otimes n}\right\rangle$ are both hard to prepare on NISQ devices, as it requires implementing the relatively deep $T$ circuit, and hard to simulate on classical computers, as it can be treated as Clifford circuits with non-stabilizer input states. However, interestingly, within the SH-VQE framework, the operator expectation values under these states can be efficiently evaluated as long as $U$ is classically tractable. Second, we want to talk about the trainability of SH-VQA. A known result is that in general, an ansatz with high expressivity may lead to low trainability \cite{holmes2022connecting}. We want to emphasize that the expressivity benchmarked under the very random settings in Fig. \ref{p2} should be understood as the achievable expressivity of the NISQ devices enhanced by Heisenberg circuits but not the actual expressivity of the ansatz within the SH-VQA framework for specific problems. Thus, SH-VQA can be understood as a general methodology for improving existing variational algorithms within which biased and trainable ansatzes can be tested. We summarize some strategies in the Appendix.

In summary, we have introduced a novel variational quantum algorithm, the SH-VQA, to efficiently extend the circuit depth of near-term noisy quantum processors. By virtually introducing relatively deep and non-local Clifford circuits, we show that the expressivity of shallow quantum circuits can be significantly enhanced, without sacrificing the fidelity. We use the XXZ model to demonstrate the workflow of SH-VQA and further demonstrate the practical values of SH-VQA by solving small molecules. Our method is directly applicable to current quantum hardware and is compatible with most existing quantum algorithms. Leveraging quantum error mitigation, our work pushes near-term quantum hardware into wide non-trivial applications.

We use the Qulacs \cite{suzuki2020qulacs} and the Qiskit \cite{cross2018ibm} packages for parts of simulations.

\bibliographystyle{unsrt}
\bibliography{ref.bib}

\begin{thebibliography}{10}

\bibitem{feynman2018simulating}
Richard~P Feynman.
\newblock Simulating physics with computers.
\newblock In {\em Feynman and computation}, pages 133--153. CRC Press, 2018.

\bibitem{arute2019quantum}
Frank Arute, Kunal Arya, Ryan Babbush, Dave Bacon, Joseph~C Bardin, Rami
  Barends, Rupak Biswas, Sergio Boixo, Fernando~GSL Brandao, David~A Buell,
  et~al.
\newblock Quantum supremacy using a programmable superconducting processor.
\newblock {\em Nature}, 574(7779):505--510, 2019.

\bibitem{zhu2022quantum}
Qingling Zhu, Sirui Cao, Fusheng Chen, Ming-Cheng Chen, Xiawei Chen, Tung-Hsun
  Chung, Hui Deng, Yajie Du, Daojin Fan, Ming Gong, et~al.
\newblock Quantum computational advantage via 60-qubit 24-cycle random circuit
  sampling.
\newblock {\em Science bulletin}, 67(3):240--245, 2022.

\bibitem{wu2021strong}
Yulin Wu, Wan-Su Bao, Sirui Cao, Fusheng Chen, Ming-Cheng Chen, Xiawei Chen,
  Tung-Hsun Chung, Hui Deng, Yajie Du, Daojin Fan, et~al.
\newblock Strong quantum computational advantage using a superconducting
  quantum processor.
\newblock {\em Physical review letters}, 127(18):180501, 2021.

\bibitem{zhong2020quantum}
Han-Sen Zhong, Hui Wang, Yu-Hao Deng, Ming-Cheng Chen, Li-Chao Peng, Yi-Han
  Luo, Jian Qin, Dian Wu, Xing Ding, Yi~Hu, et~al.
\newblock Quantum computational advantage using photons.
\newblock {\em Science}, 370(6523):1460--1463, 2020.

\bibitem{zhong2021phase}
Han-Sen Zhong, Yu-Hao Deng, Jian Qin, Hui Wang, Ming-Cheng Chen, Li-Chao Peng,
  Yi-Han Luo, Dian Wu, Si-Qiu Gong, Hao Su, et~al.
\newblock Phase-programmable gaussian boson sampling using stimulated squeezed
  light.
\newblock {\em Physical review letters}, 127(18):180502, 2021.

\bibitem{preskill2018quantum}
John Preskill.
\newblock Quantum computing in the nisq era and beyond.
\newblock {\em Quantum}, 2:79, 2018.

\bibitem{endo2021hybrid}
Suguru Endo, Zhenyu Cai, Simon~C Benjamin, and Xiao Yuan.
\newblock Hybrid quantum-classical algorithms and quantum error mitigation.
\newblock {\em Journal of the Physical Society of Japan}, 90(3):032001, 2021.

\bibitem{cerezo2021variational}
Marco Cerezo, Andrew Arrasmith, Ryan Babbush, Simon~C Benjamin, Suguru Endo,
  Keisuke Fujii, Jarrod~R McClean, Kosuke Mitarai, Xiao Yuan, Lukasz Cincio,
  et~al.
\newblock Variational quantum algorithms.
\newblock {\em Nature Reviews Physics}, 3(9):625--644, 2021.

\bibitem{peruzzo2014variational}
Alberto Peruzzo, Jarrod McClean, Peter Shadbolt, Man-Hong Yung, Xiao-Qi Zhou,
  Peter~J Love, Al{\'a}n Aspuru-Guzik, and Jeremy~L O’brien.
\newblock A variational eigenvalue solver on a photonic quantum processor.
\newblock {\em Nature communications}, 5(1):4213, 2014.

\bibitem{yuan2019theory}
Xiao Yuan, Suguru Endo, Qi~Zhao, Ying Li, and Simon~C Benjamin.
\newblock Theory of variational quantum simulation.
\newblock {\em Quantum}, 3:191, 2019.

\bibitem{farhi2014quantum}
Edward Farhi, Jeffrey Goldstone, and Sam Gutmann.
\newblock A quantum approximate optimization algorithm.
\newblock {\em arXiv preprint arXiv:1411.4028}, 2014.

\bibitem{wecker2015progress}
Dave Wecker, Matthew~B Hastings, and Matthias Troyer.
\newblock Progress towards practical quantum variational algorithms.
\newblock {\em Physical Review A}, 92(4):042303, 2015.

\bibitem{kandala2017hardware}
Abhinav Kandala, Antonio Mezzacapo, Kristan Temme, Maika Takita, Markus Brink,
  Jerry~M Chow, and Jay~M Gambetta.
\newblock Hardware-efficient variational quantum eigensolver for small
  molecules and quantum magnets.
\newblock {\em nature}, 549(7671):242--246, 2017.

\bibitem{romero2018strategies}
Jonathan Romero, Ryan Babbush, Jarrod~R McClean, Cornelius Hempel, Peter~J
  Love, and Al{\'a}n Aspuru-Guzik.
\newblock Strategies for quantum computing molecular energies using the unitary
  coupled cluster ansatz.
\newblock {\em Quantum Science and Technology}, 4(1):014008, 2018.

\bibitem{temme2017error}
Kristan Temme, Sergey Bravyi, and Jay~M Gambetta.
\newblock Error mitigation for short-depth quantum circuits.
\newblock {\em Physical review letters}, 119(18):180509, 2017.

\bibitem{li2017efficient}
Ying Li and Simon~C Benjamin.
\newblock Efficient variational quantum simulator incorporating active error
  minimization.
\newblock {\em Physical Review X}, 7(2):021050, 2017.

\bibitem{brandao2015exponential}
Fernando~GSL Brandao and Micha{\l} Horodecki.
\newblock Exponential decay of correlations implies area law.
\newblock {\em Communications in mathematical physics}, 333:761--798, 2015.

\bibitem{verstraete2008matrix}
Frank Verstraete, Valentin Murg, and J~Ignacio Cirac.
\newblock Matrix product states, projected entangled pair states, and
  variational renormalization group methods for quantum spin systems.
\newblock {\em Advances in physics}, 57(2):143--224, 2008.

\bibitem{bravyi2006lieb}
Sergey Bravyi, Matthew~B Hastings, and Frank Verstraete.
\newblock Lieb-robinson bounds and the generation of correlations and
  topological quantum order.
\newblock {\em Physical review letters}, 97(5):050401, 2006.

\bibitem{ho2019efficient}
Wen~Wei Ho and Timothy~H Hsieh.
\newblock Efficient variational simulation of non-trivial quantum states.
\newblock {\em SciPost Phys}, 6:029, 2019.

\bibitem{vidal2003entanglement}
Guifre Vidal, Jos{\'e}~Ignacio Latorre, Enrique Rico, and Alexei Kitaev.
\newblock Entanglement in quantum critical phenomena.
\newblock {\em Physical review letters}, 90(22):227902, 2003.

\bibitem{latorre2003ground}
Jos{\'e}~Ignacio Latorre, Enrique Rico, and Guifr{\'e} Vidal.
\newblock Ground state entanglement in quantum spin chains.
\newblock {\em arXiv preprint quant-ph/0304098}, 2003.

\bibitem{huang2015quantum}
Yichen Huang, Xie Chen, et~al.
\newblock Quantum circuit complexity of one-dimensional topological phases.
\newblock {\em Physical Review B}, 91(19):195143, 2015.

\bibitem{chen2010local}
Xie Chen, Zheng-Cheng Gu, and Xiao-Gang Wen.
\newblock Local unitary transformation, long-range quantum entanglement, wave
  function renormalization, and topological order.
\newblock {\em Physical review b}, 82(15):155138, 2010.

\bibitem{zhang2022variational}
Shi-Xin Zhang, Zhou-Quan Wan, Chee-Kong Lee, Chang-Yu Hsieh, Shengyu Zhang, and
  Hong Yao.
\newblock Variational quantum-neural hybrid eigensolver.
\newblock {\em Physical Review Letters}, 128(12):120502, 2022.

\bibitem{mizukami2020orbital}
Wataru Mizukami, Kosuke Mitarai, Yuya~O Nakagawa, Takahiro Yamamoto, Tennin
  Yan, and Yu-ya Ohnishi.
\newblock Orbital optimized unitary coupled cluster theory for quantum
  computer.
\newblock {\em Physical Review Research}, 2(3):033421, 2020.

\bibitem{mcclean2016theory}
Jarrod~R McClean, Jonathan Romero, Ryan Babbush, and Al{\'a}n Aspuru-Guzik.
\newblock The theory of variational hybrid quantum-classical algorithms.
\newblock {\em New Journal of Physics}, 18(2):023023, 2016.

\bibitem{liu2018entanglement}
Zi-Wen Liu, Seth Lloyd, Elton Zhu, and Huangjun Zhu.
\newblock Entanglement, quantum randomness, and complexity beyond scrambling.
\newblock {\em Journal of High Energy Physics}, 2018(7):1--62, 2018.

\bibitem{gottesman1998heisenberg}
Daniel Gottesman.
\newblock The heisenberg representation of quantum computers.
\newblock {\em arXiv preprint quant-ph/9807006}, 1998.

\bibitem{hoggar1982t}
Stuart~G Hoggar.
\newblock t-designs in projective spaces.
\newblock {\em European Journal of Combinatorics}, 3(3):233--254, 1982.

\bibitem{Note2}
Data source.
\newblock \url{https://www.polyu.edu.hk/ama/staff/xjchen/sphdesigns.html}.

\bibitem{brandao2016local}
Fernando~GSL Brandao, Aram~W Harrow, and Micha{\l} Horodecki.
\newblock Local random quantum circuits are approximate polynomial-designs.
\newblock {\em Communications in Mathematical Physics}, 346(2):397--434, 2016.

\bibitem{zhu2017multiqubit}
Huangjun Zhu.
\newblock Multiqubit clifford groups are unitary 3-designs.
\newblock {\em Physical Review A}, 96(6):062336, 2017.

\bibitem{mcclean2018barren}
Jarrod~R McClean, Sergio Boixo, Vadim~N Smelyanskiy, Ryan Babbush, and Hartmut
  Neven.
\newblock Barren plateaus in quantum neural network training landscapes.
\newblock {\em Nature communications}, 9(1):4812, 2018.

\bibitem{arouri2020accelerated}
Yazan Arouri and Mohammad Sayyafzadeh.
\newblock An accelerated gradient algorithm for well control optimization.
\newblock {\em Journal of Petroleum Science and Engineering}, 190:106872, 2020.

\bibitem{hein2006entanglement}
Marc Hein, Wolfgang D{\"u}r, Jens Eisert, Robert Raussendorf, M~Nest, and H-J
  Briegel.
\newblock Entanglement in graph states and its applications.
\newblock {\em arXiv preprint quant-ph/0602096}, 2006.

\bibitem{zhang2020differentiable}
Shi-Xin Zhang, Chang-Yu Hsieh, Shengyu Zhang, and Hong Yao.
\newblock Differentiable quantum architecture search.
\newblock {\em arXiv preprint arXiv:2010.08561}, 2020.

\bibitem{Note1}
{\em The possibility of picking a graph $\vec{k}={ }^{\prime} k_{1} k_{2}
  \ldots k_{\lfloor n / 2\rfloor}$ ' is described by the product of $\lfloor n
  / 2\rfloor$ independent softmax functions $\prod_{i=1}^{\lfloor n / 2\rfloor}
  \frac{\exp \left(\alpha_{i, k_{i}}\right)}{\exp \left(\alpha_{i,
  0}\right)+\exp \left(\alpha_{i, 1}\right)}$.}

\bibitem{mcardle2020quantum}
Sam McArdle, Suguru Endo, Al{\'a}n Aspuru-Guzik, Simon~C Benjamin, and Xiao
  Yuan.
\newblock Quantum computational chemistry.
\newblock {\em Reviews of Modern Physics}, 92(1):015003, 2020.

\bibitem{bravyi2002fermionic}
Sergey~B Bravyi and Alexei~Yu Kitaev.
\newblock Fermionic quantum computation.
\newblock {\em Annals of Physics}, 298(1):210--226, 2002.

\bibitem{holmes2022connecting}
Zo{\"e} Holmes, Kunal Sharma, Marco Cerezo, and Patrick~J Coles.
\newblock Connecting ansatz expressibility to gradient magnitudes and barren
  plateaus.
\newblock {\em PRX Quantum}, 3(1):010313, 2022.

\bibitem{suzuki2020qulacs}
Yasunari Suzuki, Yoshiaki Kawase, Yuya Masumura, Yuria Hiraga, Masahiro
  Nakadai, Jiabao Chen, Ken~M Nakanishi, Kosuke Mitarai, Ryosuke Imai, Shiro
  Tamiya, et~al.
\newblock Qulacs: a fast and versatile quantum circuit simulator for research
  purpose.
\newblock {\em arXiv preprint arXiv:2011.13524}, 2020.

\bibitem{cross2018ibm}
Andrew Cross.
\newblock The ibm q experience and qiskit open-source quantum computing
  software.
\newblock In {\em APS March meeting abstracts}, volume 2018, pages L58--003,
  2018.

\end{thebibliography}
\onecolumngrid
\appendix
\counterwithin{figure}{section} 
\section{Measurement cost}

For variational hybrid quantum-classical algorithms, a key step is to evaluate operators' expectation values. Typically, we use the so-called operator averaging method, which has no requirements on circuits but requires a large number of measurements.

Without loss of generality, we consider a simple original Hamiltonian $H$, which is composed of only one Pauli operator. Consider $P_{h}$ to be the Pauli operator we want to evaluate $\left\langle P_{h}\right\rangle=\operatorname{tr}\left(\rho P_{h}\right)$, the variance of measuring $P_{h}$ is $\operatorname{Var}\left[P_{h}\right]_{\rho}=\operatorname{tr}\left(\rho P_{h}^{2}\right)-$ $\operatorname{tr}\left(\rho P_{h}\right)^{2}$. If we repeat the measurement for $N_{1}$ times, the variance becomes

\begin{equation}\label{s1}
\operatorname{Var}\left[P_{h}\right]_{\rho} \rightarrow \frac{\operatorname{Var}\left[P_{h}\right]_{\rho}}{N_{1}}
\end{equation}

If we want to reach an accuracy of $\epsilon$, the number of measurements we need is

\begin{equation}\label{s2}
N_{1}=\frac{\operatorname{Var}\left[P_{h}\right]_{\rho}}{\epsilon^{2}}
\end{equation}

Consider we act a circuit $T$ to the Pauli operator $P_{h}$ :

\begin{equation}\label{s3}
P_{h} \rightarrow T^{\dagger} P_{h} T=\sum_{i=1}^{m_{h}} c_{i} P_{i}
\end{equation}

where $\sum_{i=1}^{m_{h}} c_{i}^{2}=1$. The circuit transforms $P_{h}$ into $m_{h}$ part. For the corresponding state $\rho_{T}=T^{\dagger} \rho T$, the variance is unchanged under the transformation $\operatorname{Var}\left[P_{h}\right]_{\rho}=$ $\operatorname{Var}\left[T^{\dagger} P_{h} T\right]_{\rho_{T}}$. However, we need to measure each term individually so the effective variance of those measurements does increase. One natural way is to repeat $N_{2} / m_{h}$ measurements to evaluate each one of Pauli terms and sum them. The variance in this way is

\begin{equation}\label{s4}
\sum_{i=1}^{m_{h}} \frac{m_{h} c_{i}^{2} \operatorname{Var}\left[P_{i}\right]_{\rho_{T}}}{N_{2}}
\end{equation}

If we want to reach the same accuracy of $\epsilon$, the number of measurements $N_{2}$ we need is

\begin{equation}\label{s5}
N_{2}=\frac{m_{h} \sum_{i=1}^{m_{h}} c_{i}^{2} \operatorname{Var}\left[P_{i}\right]_{\rho_{T}}}{\epsilon^{2}} \approx m_{h} \frac{\operatorname{Var}\left[P_{h}\right]_{\rho}}{\epsilon^{2}}=m_{h} N_{1}
\end{equation}

The approximation above is because every $\left\langle P_{i}\right\rangle$ can be treated as an expectation value of a Bernoulli random variable $\left\langle P_{i}\right\rangle=p_{1} * 1+p_{-1} *(-1)$ and the variance is bounded by $4 p_{1} p_{-1} \leq 1$. So, we assume the variances of Pauli terms are at the same level. From Eq. \ref{s5} we can see $m_{h}$ must be controlled at a tolerable level, which explains why the structure of VHC must be restricted. This conclusion does not change even if one groups commuting terms.

Another point worth mentioning is that assigning $N_{2} / m_{h}$ measurements to evaluate each one of Pauli's terms is not the best choice. One can treat this as an optimization problem

\begin{eqnarray}\label{s6}
\text { minimize: } && f\left(p_{i}\right)=\sum_{i} \frac{c_{i}^{2} \operatorname{Var}\left[P_{i}\right]_{\rho_{T}}}{N_{2} p_{i}} \nonumber\\
\text { subject to: } && \sum_{i} p_{i}=1 \text { and } p_{i} \geq 0, i=1, \ldots, m_{h}
\end{eqnarray}

This problem can be easily solved using the Lagrange multiplier method under the assumption that the variances of Pauli terms are at the same level and the best choice is $p_{i}=\left|c_{i}\right| / \sum_{i=1}^{m_{h}}\left|c_{i}\right|$ (which won't change the basic conclusion). Another choice is $p_{i}=$ $c_{i}^{2}$. At first look, this choice should be better than the uniform choice $p_{i}=1 / m_{h}$ as the term with bigger variance is assigned with more measurements. However, they actually have the same performances

\begin{equation}\label{s7}
\sum_{i=1}^{m_{h}} \frac{c_{i}^{2} \operatorname{Var}\left[P_{i}\right]_{\rho_{T}}}{N_{2} c_{i}^{2}}=\frac{\sum_{i=1}^{m_{h}} \operatorname{Var}\left[P_{i}\right]_{\rho_{T}}}{N_{2}} \approx \sum_{i=1}^{m_{h}} \frac{m_{h} c_{i}^{2} \operatorname{Var}\left[P_{i}\right]_{\rho_{T}}}{N_{2}}
\end{equation}

\section{The structure of parametrized circuit}

\begin{figure*}[htb]
\includegraphics[width=0.8\textwidth]{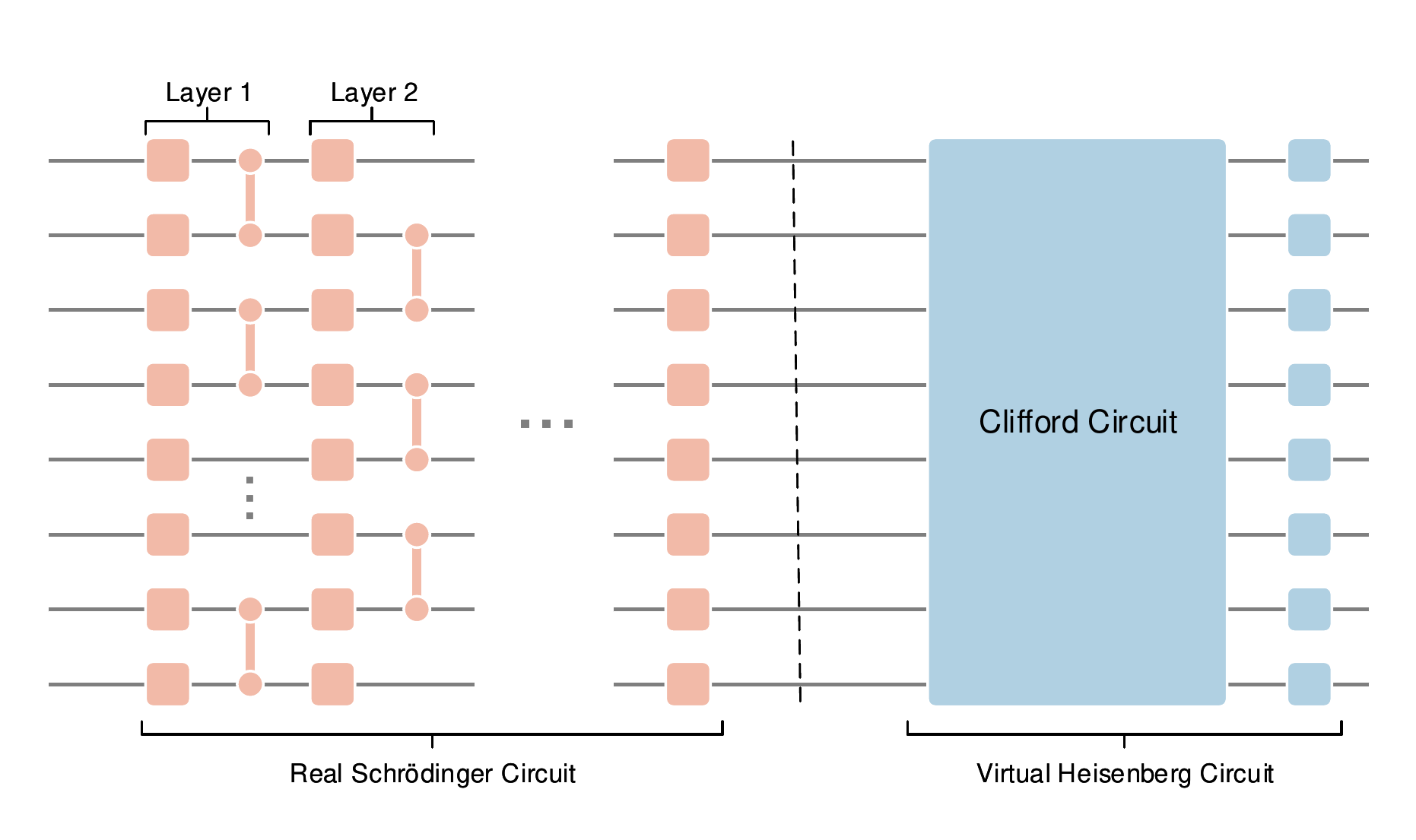}
\caption{\label{ps4} Structure of parametrized circuits used when solving XXZ models. Two qubit gates are fixed $\mathrm{CZ}$ gates while each single-qubit gate is parametrized as $\exp \left(-i \theta_{\mathrm{x}} \sigma_{\mathrm{x}}\right) \exp \left(-\mathrm{i} \theta_{\mathrm{y}} \sigma_{\mathrm{y}}\right) \exp \left(-\mathrm{i} \theta_{\mathrm{z}} \sigma_{\mathrm{z}}\right)$. The Clifford part is composed of only CZ gates.}
\end{figure*}

The structure of parametrized circuits used when solving XXZ models is shown in Fig. \ref{ps4}.

\section{The expressivity and the trainability of SH-VQA}

According to Ref. \cite{holmes2022connecting}, there is a general trade-off between the expressivity and the trainability of an ansatz. Specifically, higher expressivity leads to low variance of the cost gradients. Based on this result, we can talk about the trainability of SH-VQA. The expressivity in Fig. \ref{p2} is the highest expressivity that can be achieved by SH-VQA. If we use an ansatz with the same setting as Fig. \ref{p2} for solving problems, the trainability will be rather poor. However, in real situations, for specific problems, we can have numerous strategies to restrict expressivity. 

To make SH-VQA trainable, a good trainability Schrödinger circuit is a prerequisite such as the Unitary Coupled Cluster \cite{romero2018strategies} and the Hamiltonian Variational Ansatz \cite{wecker2015progress}. Under this condition, we can further restrict the size of the pool of Clifford circuits. Since each Clifford circuit maps the exploration subspace of $U$ to another subspace, we can think the expressivity is proportional to the number of Clifford circuits in the pool in the worst case where the mapped subspaces of different Clifford circuits are orthogonal. The number of all possible Clifford circuits is exponentially large but we can restrict a small part of it using the prior knowledge of Hamiltonians, good Schrödinger ansatzes, and NISQ devices. An example is to choose Clifford circuits that transform the connectivity of NISQ circuits to be close to those problem-inspired ansatzes like UCC and HVA with good trainability. If we can fix it or choose it from a restricted set, the $t$-design order can hardly change. There are also other strategies as introduced in Ref. \cite{holmes2022connecting}. For example, we can alternatively optimize the parameters in the Schrödinger circuit and the parameters in the Heisenberg circuits. When one part is fixed, the expressivity only depends on the other part. Note that both $U$ and $T$ are circuits with poor expressivity, it is the combination of them that has high expressivity. 

\begin{figure*}[htb]
\includegraphics[width=0.9\textwidth]{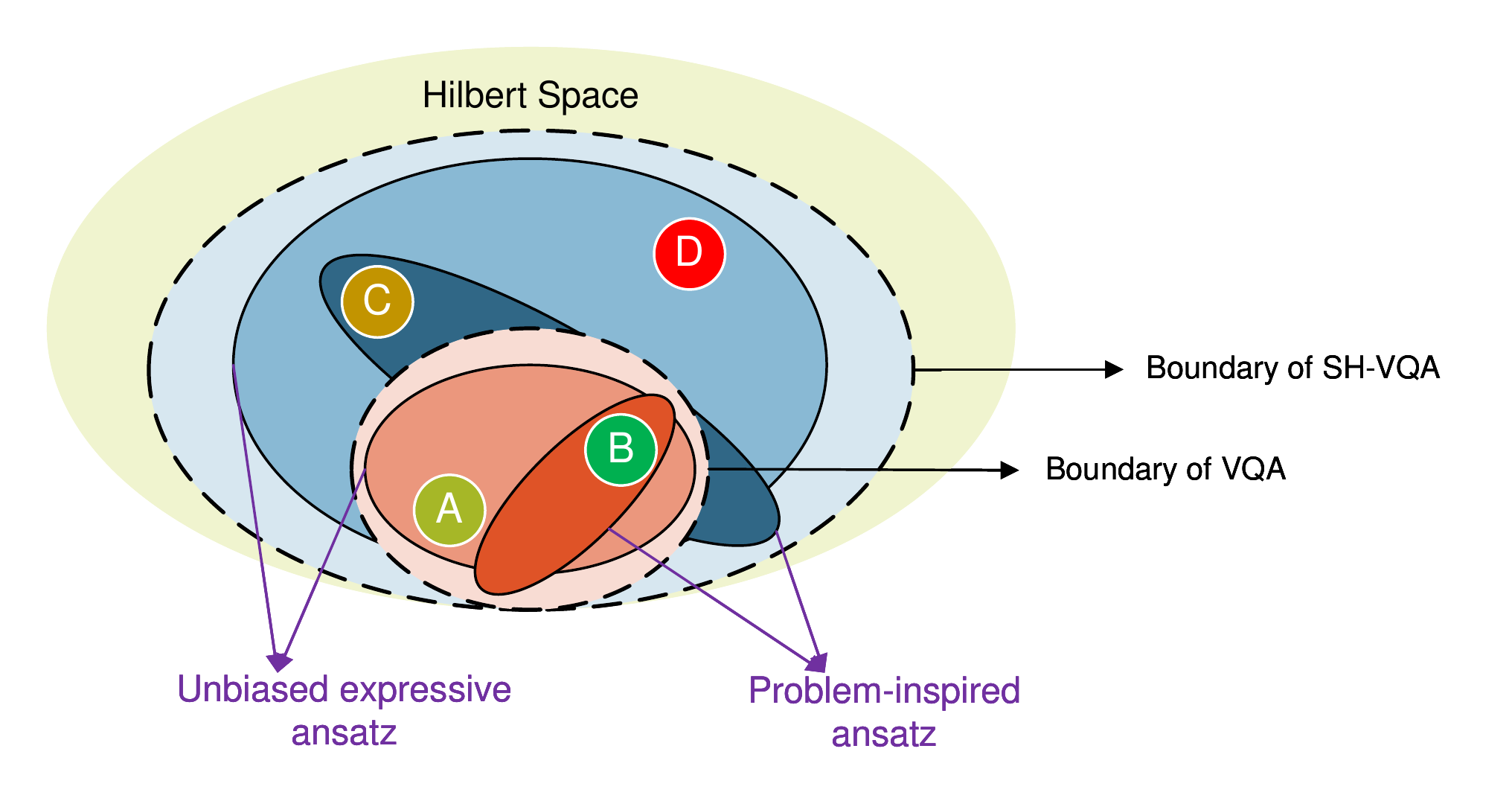}
\caption{\label{ps5} The expressivity and the trainability of SH-VQA. The motivation of our SH-VQA proposal is to go beyond the VQA boundary set by the NISQ devices and give us more freedom to build the ansatzes. SH-VQA is more of a general methodology for improving existing variational algorithms than a specific algorithm. Given specific problems $(A, B, C, D)$, we want to use the real quantum circuit to solve them. We can see that for VQA, we can only possibly solve problems $A$ and $B$ since $C$ and $D$ are outside the VQA boundary. On the contrary, we are possible to solve all A, B, C, and D by SH-VQA, which shows the advantages of our proposal. When really running the algorithms, we are required to build ansatzes. We know that for VQA, there are either unbiased ansatz with high expressivity but low trainability (e.g., the hardware-efficient ansatz) or problem-inspired ansatz with biased low expressivity and high trainability (e.g., the UCC and the HVA). Then, similar to the VQA, for SH-VQA, we can also expect to propose unbiased expressive ansatz or problem-inspired ansatz.}
\end{figure*}

The purpose of the above discussions is to show that we are able to find biased problem-inspired trainable SH-VQA ansatzes, which have no conflicts with the expressivity enhancement shown in Fig. \ref{p2}. We can have a better understanding from Fig. \ref{ps5}

\section{Circuit expressivity and hardware requirements}

\begin{figure*}[htb]
\includegraphics[width=0.9\textwidth]{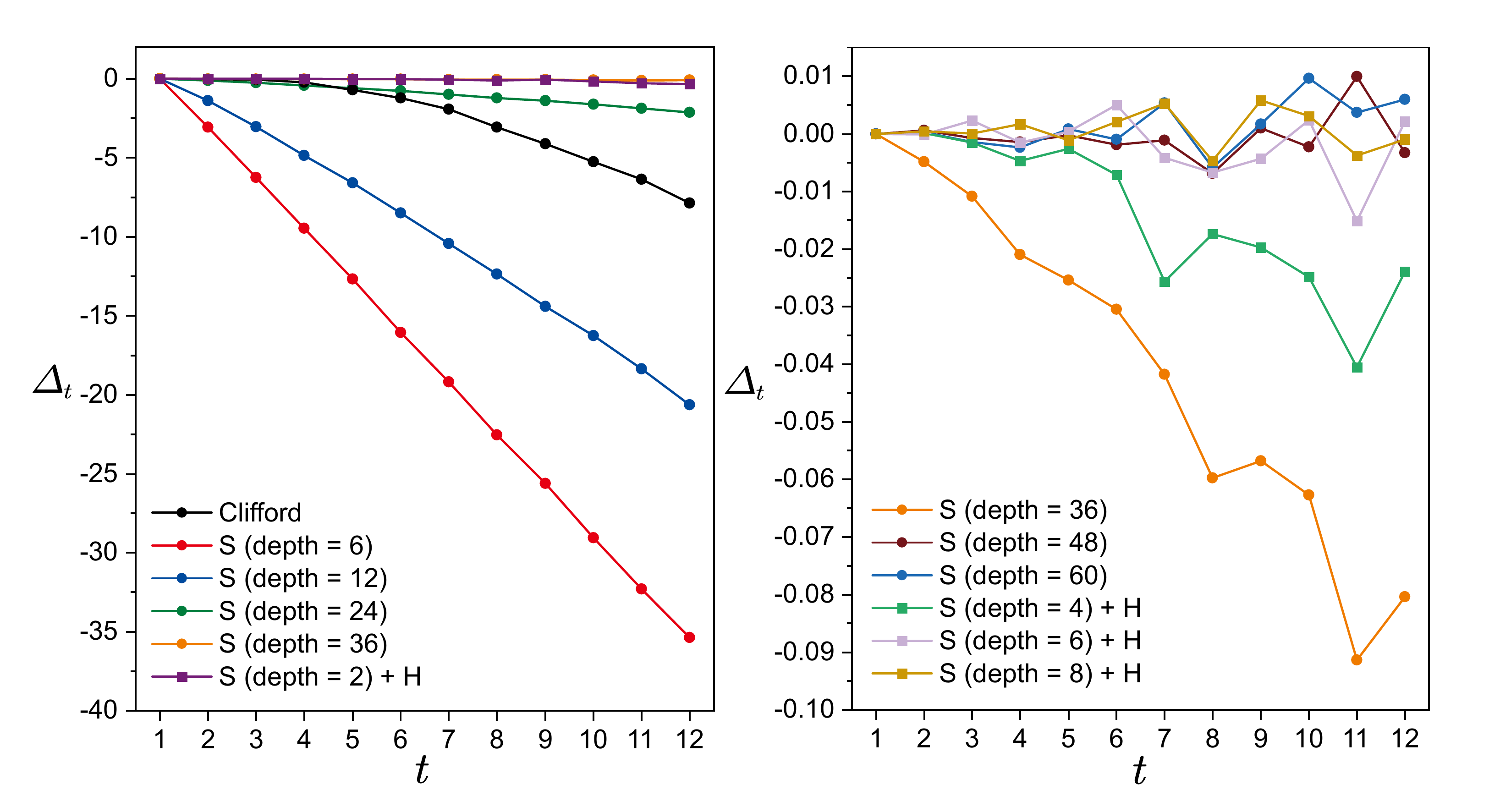}
\caption{\label{ps1} Values of 12-qubit $\Delta_{t}$ from $t=1$ to $t=12$ with different circuit settings. SH-VQE circuits of very shallow Schrödinger depth have the expressivity close to very deep VQE circuits. The Clifford circuits as a reference can generate a 3-design and a good approximation of the 4-design. SH-VQE circuits of Schrödinger depth 6 are believed to generate max scrambled states as values of $\Delta_{t}$ from $t=1$ to $t=12$ are all zero. Observing such a phenomenon in VQE circuits requires a depth of around 48. Each point is obtained by averaging 4000 samples.}
\end{figure*}

\begin{figure}[htb]
\includegraphics[width=0.6\textwidth]{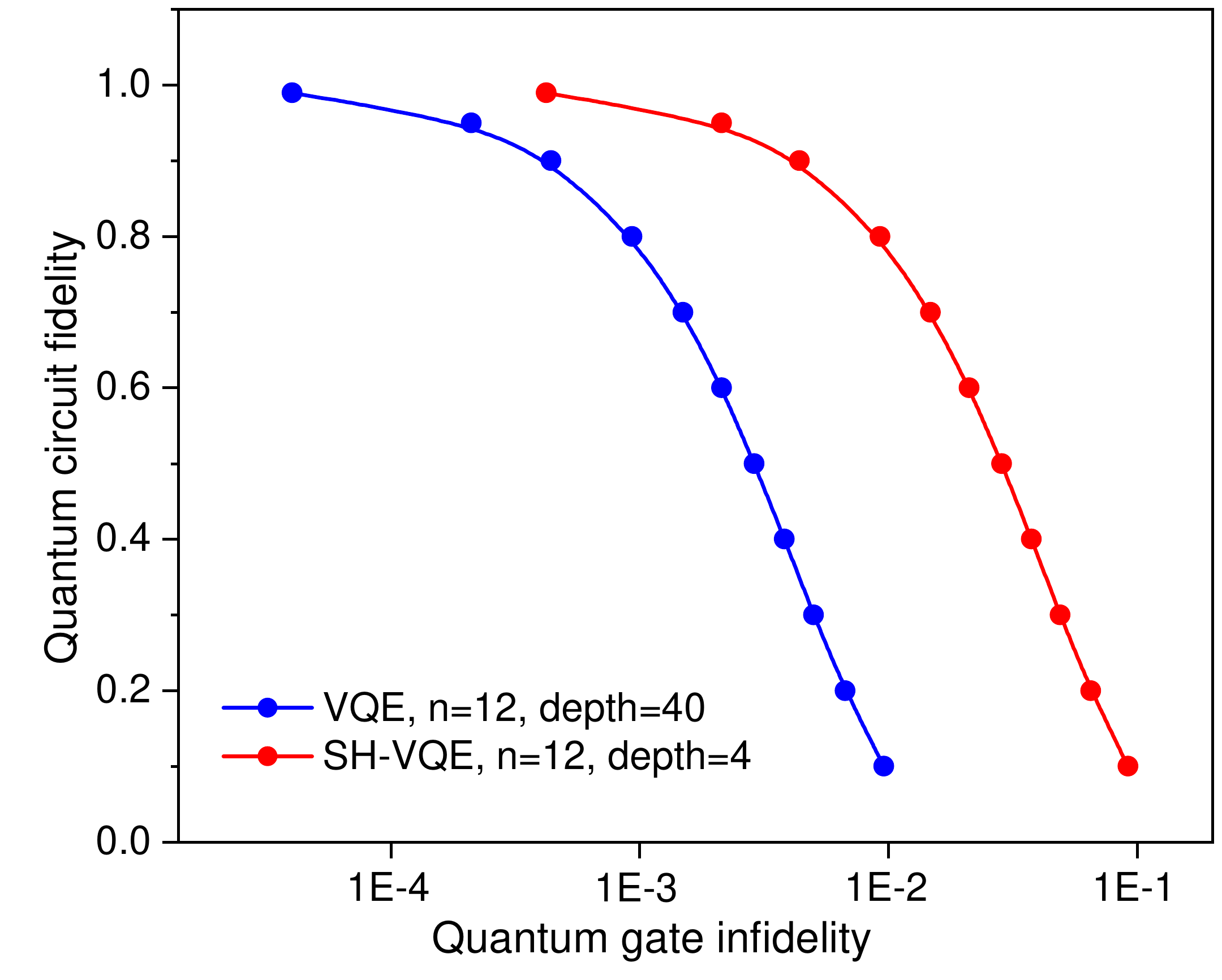}
\caption{\label{ps2} Fidelity requirement relief. The infidelity of 2-qubit blocks is plotted as a function of total circuit fidelity. The expressivity of the 12-qubit 4-depth SH-VQE circuit is close to the 12-qubit 40-depth VQE circuit whereas there is a huge difference in the fidelity requirement of gate blocks.}
\end{figure}

More detailed 12-qubit simulation results of expressivity comparisons between SH-VQE circuits and VQE circuits are shown in Fig. \ref{ps1}. From Fig. \ref{ps1}, we observe that by using the idea of SH-VQE, quantum hardware with a depth of 4 may have a close potential to 40-depth hardware running VQE. This provides us a significant relief on the requirement of quantum gate fidelities in practical experiments. We show such a reduction in Fig. \ref{ps2}. 

\begin{figure}[htb]
\includegraphics[width=0.6\textwidth]{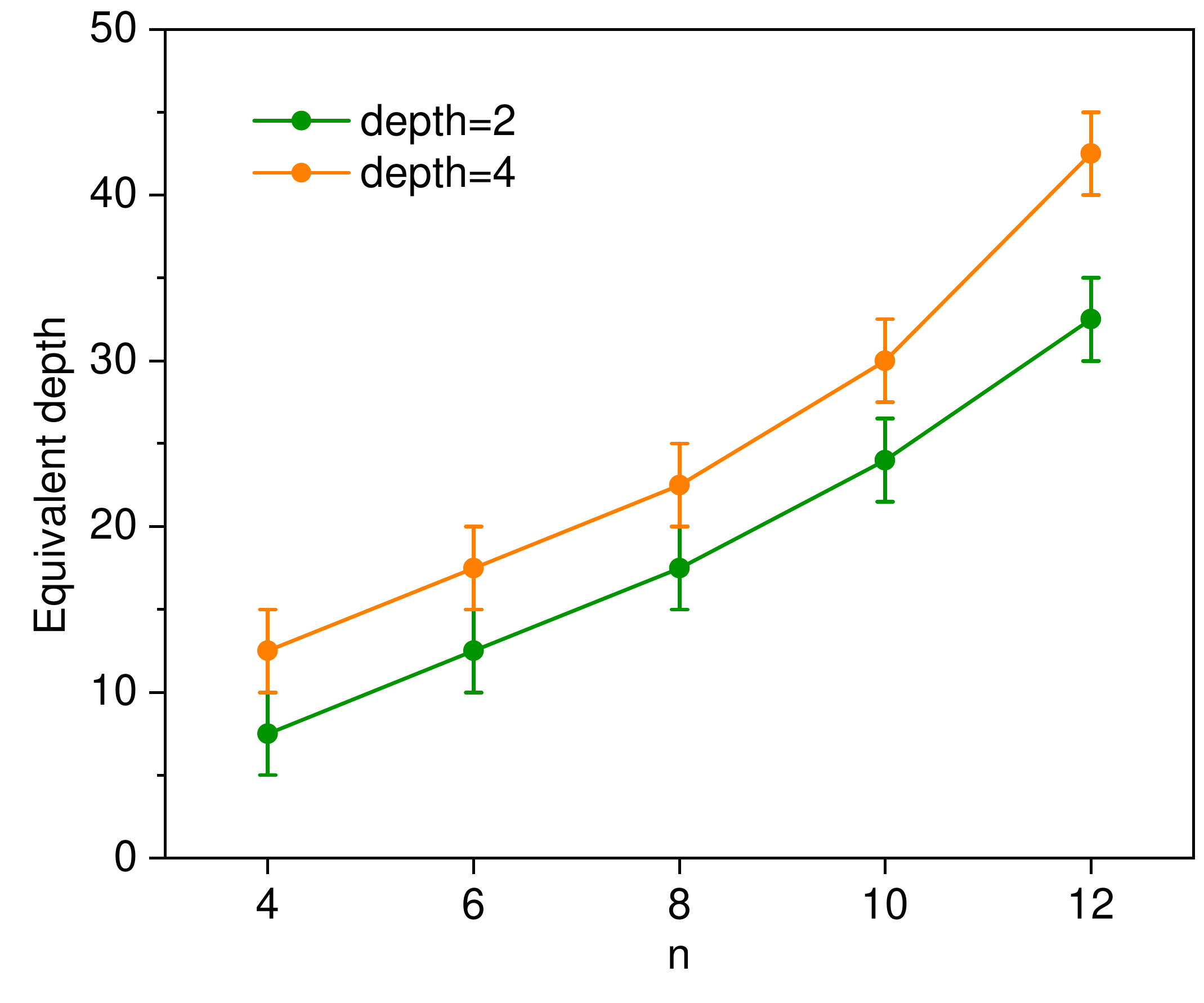}
\caption{\label{ps3} Algorithm scalability. For the SH-VQE circuit, we fix the Schrödinger circuit depth at 2 and 4 and add Clifford circuits as the Heisenberg part. For different system sizes, we show the equivalent VQE circuit depth by comparing the expressivity. The depth improvement becomes stronger and stronger as the number of qubits increases.}
\end{figure}

We further run numerical experiments to see the scalability of our algorithm. We treat expressivity as a measure of the equivalence of VQE circuit and the SH-VQE circuit. The results of our algorithm are shown in Fig. \ref{ps3}, where we fix the SH-VQE Schrödinger circuit depth at 2 and 4 and show the equivalent VQE circuit depth as functions of the system size. From the figure, we can see that the more of the qubit number is, the more equivalent VQE circuit depth is achieved, which gives SH-VQE a growing advantage as the quantum device become larger and larger. 

\end{document}